\documentclass[12pt]{article}

\usepackage[]{graphicx}
\usepackage[]{xcolor}
\makeatletter
\def\maxwidth{ %
	\ifdim\Gin@nat@width>\linewidth
	\linewidth
	\else
	\Gin@nat@width
	\fi
}
\makeatother

\definecolor{fgcolor}{rgb}{0.345, 0.345, 0.345}
\newcommand{\hlnum}[1]{\textcolor[rgb]{0.686,0.059,0.569}{#1}}%
\newcommand{\hlstr}[1]{\textcolor[rgb]{0.192,0.494,0.8}{#1}}%
\newcommand{\hlopt}[1]{\textcolor[rgb]{0,0,0}{#1}}%
\newcommand{\hlstd}[1]{\textcolor[rgb]{0.345,0.345,0.345}{#1}}%
\newcommand{\hlkwb}[1]{\textcolor[rgb]{0.69,0.353,0.396}{#1}}%
\newcommand{\hlkwc}[1]{\textcolor[rgb]{0.333,0.667,0.333}{#1}}%
\newcommand{\hlkwd}[1]{\textcolor[rgb]{0.737,0.353,0.396}{\textbf{#1}}}%

\usepackage{framed}
\makeatletter
\newenvironment{kframe}{%
 \def\at@end@of@kframe{}%
 \ifinner\ifhmode%
  \def\at@end@of@kframe{\end{minipage}}%
  \begin{minipage}{\columnwidth}%
 \fi\fi%
 \def\FrameCommand##1{\hskip\@totalleftmargin \hskip-\fboxsep
 \colorbox{shadecolor}{##1}\hskip-\fboxsep
     \hskip-\linewidth \hskip-\@totalleftmargin \hskip\columnwidth}%
 \MakeFramed {\advance\hsize-\width
   \@totalleftmargin\z@ \linewidth\hsize
   \@setminipage}}%
 {\par\unskip\endMakeFramed%
 \at@end@of@kframe}
\makeatother

\definecolor{shadecolor}{rgb}{.97, .97, .97}
\definecolor{messagecolor}{rgb}{0, 0, 0}
\definecolor{warningcolor}{rgb}{1, 0, 1}
\definecolor{errorcolor}{rgb}{1, 0, 0}
\newenvironment{knitrout}{}{} % an empty environment to be redefined in TeX

\usepackage{alltt}
\usepackage{graphicx}
\usepackage{a4wide}
\usepackage[linesnumbered, longend ]{algorithm2e}
\usepackage{amssymb, amsmath, amsthm}

\usepackage[english]{babel}
\usepackage{enumitem}
\usepackage{float}
\usepackage{epstopdf}
\usepackage{pgf, tikz}
\usepackage{subcaption}
\usepackage{url}
\usepackage{xcolor}
\usepackage{booktabs}
\usepackage{thmtools}

\usepackage[colorlinks, 
linkcolor=red!50!black, 
citecolor=green!50!black, 
urlcolor=blue!50!black]{hyperref}
\usepackage[utf8]{inputenc}
\usepackage[super, sort&compress]{natbib}
\usepackage[capitalise,noabbrev,english]{cleveref}

\newcommand{\R}{\mathbb R}
\renewcommand{\P}{\mathbb{P}}

\newcommand{\eps}{\varepsilon}
\renewcommand{\phi}{\varphi}
\newcommand{\e}{\mathrm{e}}

\newcommand{\nd}{\mathcal{N}}

\DeclareMathOperator{\E}{\mathbb{E}}

\IfFileExists{upquote.sty}{\usepackage{upquote}}{}

\begin{document}
%\SweaveOpts{width=70}

%	\title{A brief introduction on latent variable based ordinal regression models with an application to survey data}
	
\title{\vspace{-20mm}A brief introduction on latent variable based ordinal regression models with an application to survey data}

\author{Johannes Wieditz\thanks{Corresponding author: Johannes Wieditz, Department of Medical Statistics, University Medical Centre Göttingen, Humboldtallee 32, 37073 Göttingen, Germany. Phone: +49 551 / 39 63029. Email: \texttt{\href{mailto:johannes.wieditz@med.uni-goettingen.de}{johannes.wieditz@med.uni-goettingen.de}}}\,\,\textsuperscript{,}\hspace{-0pt}\thanks{Department of Medical Statistics, University Medical Centre Göttingen, Humboldtallee 32, 37073~Göttingen, Germany.}\,\,\textsuperscript{,}\hspace{-0pt}\thanks{Department of Anaesthesiology, University Medical Centre Göttingen, Robert-Koch-Straße 40, 37075~Göttingen, Germany.} \and
		Clemens Miller\thanks{Neurosurgical Critical Care Unit, Department of Neurosurgery, Medical University Innsbruck, Anichstrasse 35, 6020~Innsbruck, Austria} \and Jan Scholand\footnotemark[3] \and Marcus Nemeth\footnotemark[3]
	}

\maketitle

	\textbf{Abstract: } The analysis of survey data is a frequently arising issue in clinical trials, particularly when capturing quantities which are difficult to measure. Typical examples are questionnaires about patient's well-being, pain, or consent to an intervention. In these, data is captured on a discrete scale containing only a limited number of possible answers, from which the respondent has to pick the answer which fits best his/her personal opinion. This data is generally located on an ordinal scale as answers can usually be arranged in an ascending order, e.g., ``bad", ``neutral", ``good" for well-being. 
	
	Since responses are usually stored numerically for data processing purposes, analysis of survey data using ordinary linear regression models are commonly applied. However, assumptions of these models are often not met as linear regression requires a constant variability of the response variable and can yield predictions out of the range of response categories. By using linear models, one only gains insights about the mean response which may affect representativeness.
	
	In contrast, ordinal regression models can provide probability estimates for all response categories and yield information about the full response scale beyond the mean. In this work, we provide a concise overview of the fundamentals of latent variable based ordinal models, applications to a real data set, and outline the use of state-of-the-art-software for this purpose. Moreover, we discuss strengths, limitations and typical pitfalls. This is a companion work to a current vignette-based structured interview study in paediatric anaesthesia.

	\medskip
	
	\noindent
	\textit{Keywords:} Ordinal regression, latent variable, cumulative link models, logistic regression, response distribution, factors influencing willingness to consent to participation

	\section{Introduction}
	
	Data derived from surveys or patient interviews are often subject of research in medicine. As answers are usually encoded numerically, e.g., using the numeric rating scale for pain, data analysis using linear models often seems reasonable. Whereas for finely graduated response scales summary statistics such as mean or median response are often of interest, these are often not very meaningful or representative for response scales with only a few categories, e.g., ``On a scale of 1 (absolutely yes) to 5 (absolutely no), do you consent in the participation in the following study?''. In this case, probabilities for the individual response levels or the proportion of responders who at least ``rather consent'' for participation would be more revealing. The application of this statistical methodology is illustrated using a vignette-based interview study from paediatric anaesthesia (German Clinical Trials Register \href{https://drks.de/search/de/trial/DRKS00027090}{DRKS00027090}).
	
	\subsection{Motivation}
	\label{sec:motivation}
	
	The setting of the study is as follows: Parents or legal representatives of inpatient children were asked for their willingness to consent to participation in three \textit{fictional} studies containing potential objectives for clinical studies in paediatric anaesthesia that widely differed in terms of invasiveness. Aim of this investigation was to identify factors influencing the willingness to participate (acronym \textsc{Filippa}) in the following studies:
	\begin{enumerate}[label=(\arabic*)]
		\item a prospective \textit{observational study} on a non-invasive temperature measurement sensor,
		\item a \textit{randomised controlled trial} (RCT) of inducing anaesthesia by intravenous vs.\ inhalative agents, and
		\item a \textit{pharmacological study} on intravenous ibuprofen (painkiller) designed to collect data for regular approval in children, corresponding to an open label phase-II pharmacological study.
	\end{enumerate}
	Interviews were conducted by the same investigator to ensure consistent wording and in the same order (defined by increasing invasiveness as outlined above). Legal representatives were asked if they were willing to have their child participating in the corresponding studies. Responses were captured on a five-point Likert-scale of ``\textit{absolutely consent}", ``\textit{rather consent}", ``\textit{unsure}", ``\textit{rather decline}" and ``\textit{absolutely decline}" participation. For further details on experimental design and the exact descriptions of the studies, we refer to Miller et al.\cite{Miller2023}. %The dataset corresponding to this study can be found at \cite{Miller2023a}.
	
\begin{figure}[h!]
		\includegraphics[width=\linewidth]{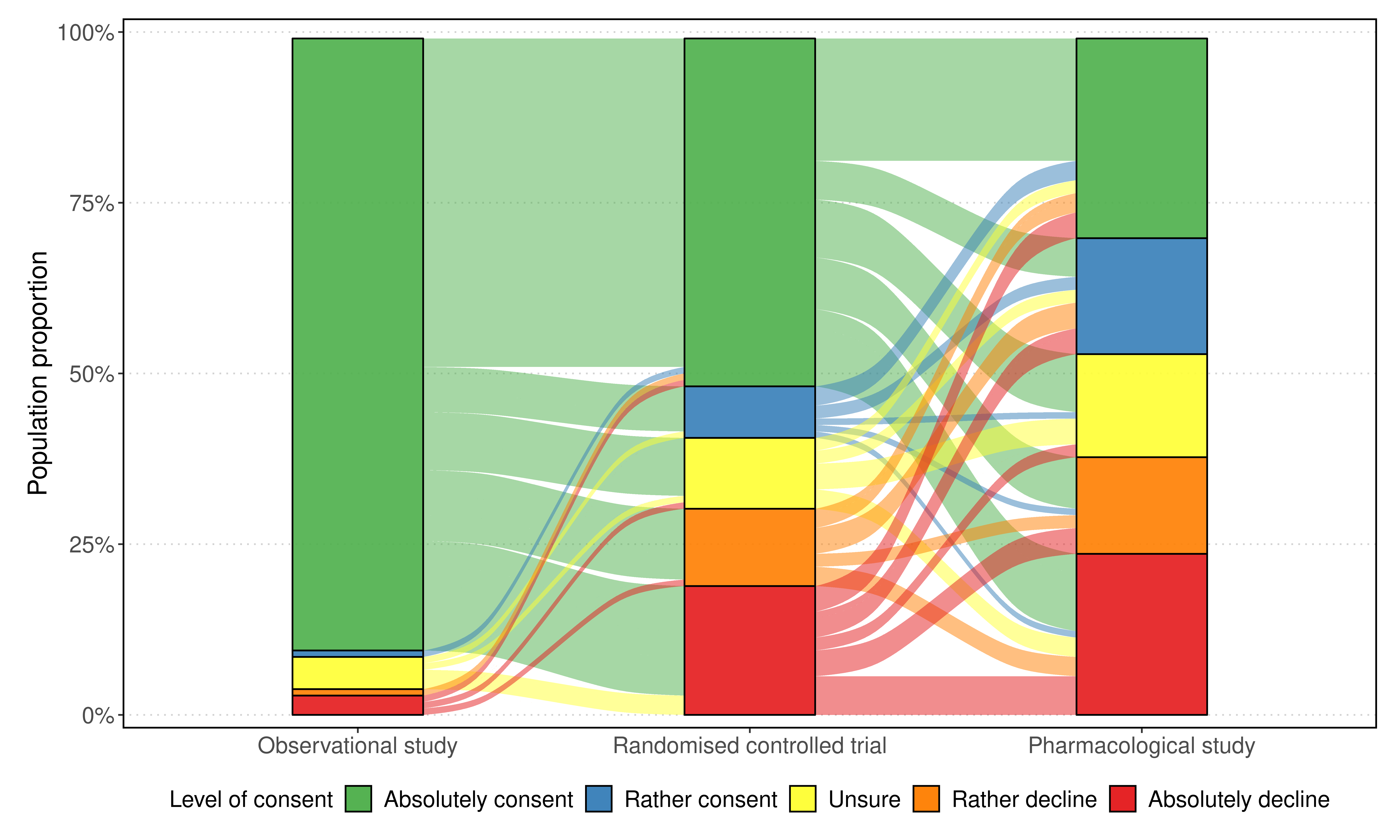}
		\caption{Alluvial diagram of the migration of the level of consent for three studies with increasing level of invasiveness  (left to right) indicated by the streams between the bars. The bar charts show the frequency distribution of the corresponding responses for a given study.}
		\label{fig:alluvial}
	\end{figure}

	\cref{fig:alluvial} portrays a descriptive statistic of the response distribution stratified by the three studies with increasing levels of invasiveness (bar charts, left to right) in form of an alluvial diagram. The streams between the bar charts show the migration between the answers from one question to the following one. It appears that as the study becomes more invasive, willingness to participate generally decreases. Note that this behaviour is, however, not present in all participants. This might be due to the fact that a certain understanding of research in medicine is beneficial to understand the different degrees of invasiveness between the studies, particularly the RCT and the pharmacological study.
	
	Statistical inference, particularly identification of significantly influencing factors or testing, however, is only possible within a statistical model. For this purpose, we perform an \textit{ordinal regression} on the response depending on the study type and including various covariates such as child's sex, age, preceding participation in studies, as well as age and professional medical degree of the legal representatives. 
	
	The applied statistical approach presented here is based on the literature briefly summarised below. Agresti\cite{Agresti2012} provides a comprehensive introduction to categorial data analysis but early work can already be found in Hildebrand et al.\cite{Hildebrand1977} and Johnson \& Albert\cite{Johnson2006}. Agresti\cite{Agresti2013} and McCullagh\cite{McCullagh1980} focus particularly on ordinal models, and Fullerton\cite{Fullerton2009} provides a concise comparison of different types of models. Goodness of fit and model selection problems as well as summary measures of the model's predictive power are addressed in Fagerland \& Hosmer; Agresti \& Kateri; Agresti \& Tarantola \cite{Fagerland2016,Agresti2017,Agresti2018}. Most ordinal regression fall into the category of \textit{generalised linear models}. To this end, Agresti \cite{Agresti2015} provide an detailed derivation. A software implementation based on these generalised linear models (so-called \textit{cumulative link models}) as well as a comprehensive how-to-apply-it tutorial is given in Christensen\cite{Christensen2015,Christensen2018}. A clear online introduction using software examples can also be found under Dunn\cite{Dunn2020,Dunn2020a}; the latter also covers issues from Bayesian ordinal regression. A broad overview about the topic and the \texttt{rms} framework including online resources is, moreover, also provided by Harrell Jr.\cite{HarrellJr2015, HarrellJr2023a, HarrellJr2023c, HarrellJr2023, HarrellJr2023b}. More recent articles demonstrate the applications of cumulative link models in deep learning classification algorithms, see e.g.\ Vargas et al.\cite{Vargas2020}. Regarding applications in medical research, Norris et al.\cite{Norris2006} present a concise comparison between different approaches. For more finely graduated responses scales, as it is often the case for e.g.\ verbal, visual analogue or numeric scales, Heller et al.\cite{Heller2016} present an approach for data analysis. Manuguerra et al.\cite{Manuguerra2020} provide corresponding software to do so. 
	
	\subsection{Audience}

  The target audience of this article is statisticians and medical researchers with a profound quantitative background who have not applied ordinal methods so far. We provide a brief introduction to the fundamentals of latent variable based ordinal models—an approach which yields an easy starting point to the topic in our opinion. This article particularly focuses on the analysis of data on response scales with only few levels and provides practical recommendations, enriched with corresponding \texttt{R} code sections. For additional information such as alternative approaches, software or model extensions we provide references to the literature at appropriate positions.

	\subsection{\texttt{R} package \texttt{ordinal}}
	\label{sec:ordinal}
	
	For the analyses within this tutorial we employ the \texttt{R} package \texttt{ordinal}\cite{Christensen2015} which is available on CRAN, or via \url{https://github.com/runehaubo/ordinal}. Code sections are presented at suitable positions and were executed using \texttt{R} version 4.3.2\cite{RCT2023}. The package supports fitting of ordinal fixed effects models (\texttt{clm}) as well as mixed effects models (\texttt{clmm}). To use the package, no extraordinary data preprocessing is required. An overview of the data set is provided below:

\begin{knitrout}
\definecolor{shadecolor}{rgb}{0.969, 0.969, 0.969}\color{fgcolor}\begin{kframe}
\begin{alltt}
\hlstd{data} \hlopt{%>%} \hlkwd{glimpse}\hlstd{()}
\end{alltt}
\begin{verbatim}
## Rows: 318
## Columns: 8
## $ id           <fct> 2, 2, 2, 3, 3, 3, 4, 4, 4, 5, 5, 5, 6, 6, 6, 7,~
## $ age          <dbl> 1.6438356, 1.6438356, 1.6438356, 13.3315068, 13~
## $ sex          <fct> Female, Female, Female, Female, Female, Female,~
## $ partner      <chr> "Mother", "Mother", "Mother", "Father", "Father~
## $ partner_age  <dbl> 35, 35, 35, 53, 53, 53, 38, 38, 38, 34, 34, 34,~
## $ prev_studies <lgl> FALSE, FALSE, FALSE, FALSE, FALSE, FALSE, FALSE~
## $ study        <fct> Observational study, Randomised controlled tria~
## $ response     <ord> Absolutely consent, Rather consent, Rather cons~
\end{verbatim}
\end{kframe}
\end{knitrout}
	
	Presented results are based moreover on the packages \texttt{emmeans}, \texttt{ggstream} and \texttt{tidyverse}, \cite{Lenth2022,Sjoberg2021,Wickham2019}. Several graphical parameters have been outsourced; code is provided in \cref{sec:appendix}.
	
\subsection{Further reading}

There is a number of alternative approaches for ordinal regression and software packages providing tools to this end beyond the latent variable based approach presented here. The introduction of these approaches and software is out of the scope of this paper. We state some reading recommendations in the following.

Yee \cite{Yee2015} present vector generalized linear and additive models requiring only that the regression coefficients enter through a set of linear predictors. The corresponding \texttt{R} package \texttt{VGAM} can be found in Yee \cite{Yee2023}. This package is highly flexible and allows, among other things, for multivariate ordinal responses or relaxing the proportional odds assumptions in considering partial proportional odds models.

Moreover, Harrell Jr.\cite{HarrellJr2015} provides with the corresponding \texttt{R} package \texttt{rms} \cite{HarrellJr2023} a large number of functions for fitting models, estimation, tests, confidence intervals, interpreting and displaying results. To be emphasised should be the functions \texttt{orm} and \texttt{lrm} which are primarily used for continuous and discrete ordinal data, respectively. These functions also implement cumulative link models, but without scale effects, partial or structured thresholds. A Bayesian alternative is implemented in the \texttt{brms} package \cite{Buerkner2017} or by the \texttt{blrm} function of the \texttt{rmsb} package \cite{HarrellJr2023b} which often promises better convergence behaviours in presence of random effects in the model. For many additional resources for ordinal models, we refer to Harrell Jr.\cite{HarrellJr2023a, HarrellJr2023c}.

	\subsection{Outline}
	
	This article is structured as follows: \cref{sec:strengths} puts ordinary linear regression and ordinal models in juxtaposition and points out each strengths and limitations. \cref{sec:ordinal_models} introduces the fundamental ideas of latent variable based ordinal regression models. \cref{sec:discrete_covariates} considers the influence of discrete and continuous covariates on the response outcome and examine different aspects on estimated parameters and confidence intervals, prediction, goodness of fit and outline numerical issues that may arise. We allude to advanced topics in \cref{sec:advanced_topics} and summarise our findings in \cref{sec:conclusion}. All computations are stated along with corresponding \texttt{R} code sections. An interactive version of this manuscript can be found at \url{https://jwieditz.github.io/FILIPPA}.
	
	\section{Ordinal models: strengths, limitations and alternatives}
	\label{sec:strengths}
	The question of how to evaluate ordinal response data appropriately has been widely discussed in the literature, cf.~Agresti\cite[Chapter 1]{Agresti2012}. So far, however, no broad consensus has been found as the chosen approach often depends on the original question to be answered.
	
	Approaches considering ordinal responses only as mere categorical data do not exploit the additional structure. In contrast to these nominal approaches, ordinal models can provide descriptive statistics similar to ordinary linear regression, such as means, slopes or correlations. Furthermore, ordinal analysis can use a greater variety of models. These models are usually more efficient yielding higher power for detecting trends or location alternatives using fewer parameters. Moreover, these parameters are often simpler in their interpretation than parameters in standard models for nominal variables, cf.~Agresti\cite[Section 1.2]{Agresti2012}.
	
	For ordinal data with many response levels, as for instance for visual analogue/ numeric scales, ordinal models are often inappropriate as individual response levels are not that meaningful or fitting might even be infeasible if the number of parameters to be estimated grows too large\cite{Heller2016}. For this kind of data, responses are commonly already encoded numerically on the questionnaire. As a result, ordinary least squares analysis is usually less problematic and provides more interpretable insights than an ordinal analysis, cf.~Agresti\cite[Section~1.2]{Agresti2012}. 
	
	In contrast, for ordinal data consisting of only a few response levels, ordinal regression analysis is often the better choice even though a linear regression analysis can be useful for identifying variables that clearly affect the response variable. Agresti\cite[Section~1.3.2]{Agresti2012} points out a number of reasons why ordinary linear regression is in this case often inappropriate:
	
	\begin{enumerate}[label=(\alph*)]
		\item There is usually \textit{not} a clear-cut choice for the scores, i.e.\ a particular response outcome might be consistent with a range of values of some underlying latent score, modelling an abstract quantity causing the response, see \cref{sec:ordinal_models}. Ordinary regression analysis, however, does not allow for such an error.
		\item An ordinary regression approach does \textit{not} provide probability estimates for the response levels but only a prediction of the estimated value given covariates (possibly with corresponding confidence interval to quantify uncertainty).
		\item Linear regression may yield predictions beyond the original response scale (i.e.\ above the highest or below the lowest level).
		\item Linear regression ignores different variabilities in the response categories: usually there is only a small variability at predictor values for which observations fall mainly in the highest (or lowest) category, but there is a considerable variability at predictor values for which observations tend to spread among the categories.
	\end{enumerate}

	As for the presented application, see \cref{sec:motivation}, the distribution of consent and the number of people responding ``rather consent'' was of particular interest, we argue that an ordinal analysis is the most appropriate in this case. For further applications and discussions about model choices we refer to Agresti\cite[Section 1.3]{Agresti2012} and the references therein.
		
%	The primary strength of the model to highlight is that response probabilities directly corresponding to (actually existing) response probabilities of the original survey in contrast to ordinary linear models where the estimated mean response score is in general not an element of the response scale. Moreover, using the latent variable we have an easy means to compare entire response distributions of different groups in contrast to comparing just the mean response scores, yielding a more differentiated tool for comparisons.
%
%	This model flexibility, however, comes at the price of a larger number of parameters to be estimated which depends not only on the number of covariates but also on the number of response possibilities available in the questionnaire. Thus, particularly in small studies, a comprehensive ordinal model (including a lot of covariables) might not be feasible as this easily leads to overparametrized models in which not all parameters are estimable.
	
	\section{Ordinal regression models}
	\label{sec:ordinal_models}
	
	In ordinal statistics, the quantity of interest is typically an ordinal response (e.g., of a survey), modelled by a random variable $Y$ which is assumed to take value on an \textit{ordinal scale} with $L\ge 2$ levels, $1 \le 2 \le \dots \le L$, (e.g., the levels of consent). Note that although the levels are encoded numerically, it is neither assumed that we can interpret between-level distances (e.g., $2-1$) nor that the distances between two levels are equidistant---there might be a large difference between ``unsure'' and ``rather consent'' but only a small one between ``rather consent'' and ``absolutely consent''.
	
	\subsection{Latent variable approach}
	\label{sec:latent_variable_models}
	
\begin{figure}[h!]
		\includegraphics[width=\linewidth]{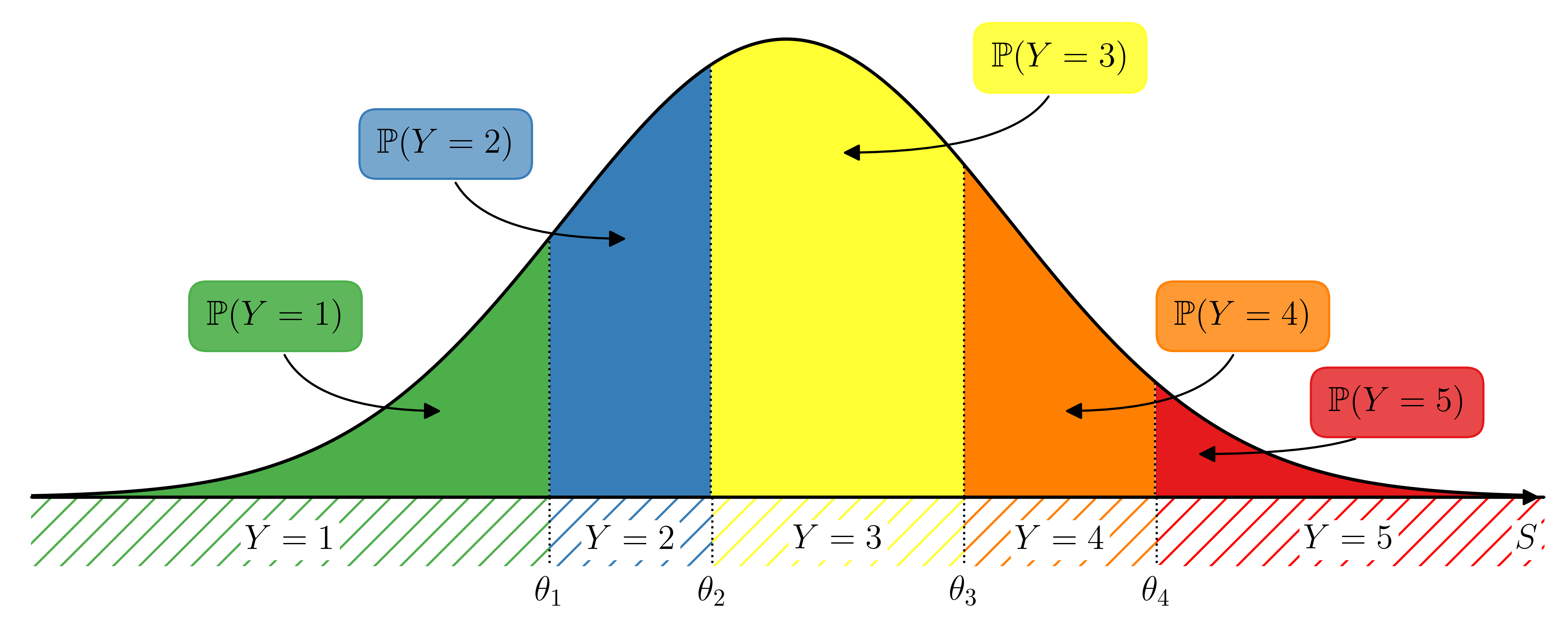}
		\caption{Probability distribution of response $Y$ (hatched regions) to a questionnaire with $L=5$ possible answers (encoded as $1$ to $5$). The response probabilities are acquired as the area (coloured regions) under the density curve of $S$ between two consecutive cut points $\theta_\ell, \theta_{\ell+1}$, $\ell=0,1,2,3,4$.}
		\label{fig:latent_variable}
	\end{figure}

	For ordinal regression, we assume that $Y$ can be acquired as the discretisation of an unobserved, continuous \textit{latent score} $S$ as
	\begin{align}
		\label{eq:defY}
		Y = \ell \ \quad \text{if and only if} \quad \theta_{\ell-1} < S \le \theta_{\ell}
	\end{align}
	for all levels $\ell = 1,2,\dots, L$ where the $\theta_\ell$'s are \textit{cut-points} (also \textit{threshold coefficients}) corresponding to the level boundaries on the scale of the latent variable, see \cref{fig:latent_variable}. The cut-points $\theta_\ell$ are assumed to be ordered in a strictly increasing manner $-\infty = \theta_0 < \theta_1 <\dots < \theta_{L-1} < \theta_L = +\infty$ and have to be estimated within a regression framework. For an ordinal variable with $L$ levels we have to estimate $L-1$ cut-points $\theta_1,\theta_2,\dots,\theta_{L-1}$. Of note, the term \textit{cut-points} is a traditional term for an endpoint in one inequality of (\ref{eq:defY}); it is not meant to suggest cutting data and is not related to categorisation or loss of information.
	
	The latent variable includes all external parameters which can influence the response behaviour of the responder and is often thought of as an abstract quantity, e.g., consent, quality of life or pain, of which $Y=1,2,\dots,L$ represent the ordinal levels e.g., ``absolutely consent", ``rather consent", ``unsure", ``rather decline", ``absolutely decline" (for $L=5$).	
	
	From \cref{eq:defY} follows that the probability distributions of $Y$ and $S$ are related as $\P(Y \le \ell) = \P(S \le \theta_{\ell})$ and in particular $\P(Y = \ell) = \P( \theta_{\ell-1} < S \le \theta_{\ell}) = \P(S \le \theta_\ell) - \P( S \le \theta_{\ell-1})$. Within an ordinal regression framework, the influence of the covariates $\textbf{x}=(x_1,x_2,\dots,x_K)^\top \in \R^K$ on the response $Y$ is typically modelled on the latent scale as
	\begin{align}
		\label{eq:nd_eps}
		S = \textbf{x}^\top \boldsymbol{\beta} + \eps = \sum_{k=1}^K x_k \beta_k + \eps
	\end{align}
	where $\boldsymbol{\beta}=(\beta_1,\beta_2,\dots,\beta_K)^\top\in\R^K$ contains the regression parameters and $\eps$ is a random variable whose distribution needs to be specified, typically with zero mean and known variance. Some common choices are stated below in \cref{ex:typical_choices}.

	\subsection{Methodological details}
	
	  \subsubsection{Link functions}
		\label{ex:typical_choices}
		\begin{enumerate}[label=(\alph*)]
			\item Assume $\eps\sim\nd(0,1)$ to be standard normally distributed. Then, we obtain the probability distribution of the survey response $Y$ given covariates $\textbf{x}$ via $S$ as
				\begin{align}
					\label{eq:prob}
					\P(Y \le \ell \mid \textbf{x}) = \P\left( S \le \theta_{\ell} \mid \textbf{x} \right) = \P\big({S - \textbf{x}^\top \boldsymbol{\beta}} \le \theta_\ell - \textbf{x}^\top \boldsymbol{\beta} \mid \textbf{x} \big) \stackrel{(\ref{eq:nd_eps})}{=} \Phi\left(\theta_\ell - \textbf{x}^\top \boldsymbol{\beta}\right),
				\end{align}
			where $\Phi$ is the distribution function of the standard normal distribution. Particularly, the probability for a response answer can be obtained as the area under a normal curve between two consecutive cut points (possibly shifted by the term including information about covariates), see \cref{fig:latent_variable}. This model is, in relation to the probit model for binary responses, also called \textit{ordered probit} model.
			\item Another popular choice is to assume $\eps$ to follow a standard logistic distribution, i.e., $\eps$~has cumulative distribution function $F(t) = \P(\eps \le t) = 1 \slash (1 + \e^{-t})$. Then, similarly to \cref{eq:prob}, we obtain
				\begin{align*}
					\P(Y \le \ell \mid \textbf{x}) = F\left(\theta_\ell - \textbf{x}^\top \boldsymbol{\beta}\right) = \frac{1}{1+\e^{-\left(\theta_\ell - \textbf{x}^\top \boldsymbol{\beta}\right)}}
				\end{align*}
			and for the log-odds holds
			\begin{align}
				\label{eq:logodds}
				\log\left(\frac{\P(Y\le \ell \mid \textbf{x})}{1-\P(Y\le \ell \mid \textbf{x})}\right) = \theta_\ell - \textbf{x}^\top \boldsymbol{\beta}, \quad \ell=1,2,\dots, L.
			\end{align}
			Thus this model is also called \textit{proportional odds} or \textit{ordered logistic regression} model \cite{McCullagh1980, Fullerton2009}. Note that for $L=2$, i.e.\ for binary responses, this model reduces to the ordinary logistic regression model.
			\item More generally, $S$ can be modelled to follow an arbitrary distribution. Christensen\cite[Section 2.2]{Christensen2018} provide a comprehensive overview about the choice of common link functions beyond the ones presented above, depending on the assumptions of the rating behaviour of the responder, e.g., choose a Cauchy distribution if extreme ratings are assumed to be more likely. Moreover, Agresti\cite[Chapter 5]{Agresti2012} yield a differentiated view on the field of ordinal models with various examples and highlight aspects about practical and theoretical issues.
		\end{enumerate}

	\subsubsection{Cumulative link models}
		\label{rem:clm}
		From \cref{eq:prob} follows (for an arbitrary choice of $\eps$) that we can write the cumulative distribution function (CDF) of $Y$ given the covariates~$\textbf{x}$ as 
		\begin{align}
			\label{eq:clm}
			\P(Y\le \ell \mid \textbf{x}) = F(\theta_\ell - \textbf{x}^\top \boldsymbol{\beta}), \quad \ell = 1,2,\dots,L,
		\end{align}
		where $F$ is the distribution function of $\eps$. As the function $F$ links the CDFs of $Y$ and $S$, ordinal models of the form (\ref{eq:clm}) are called \textit{cumulative link models} and $F$ is called \textit{inverse link function} (as it takes the linear predictor $\theta_\ell-\textbf{x}^\top \boldsymbol{\beta}$ from the latent space back and maps it to predicted probabilities for $Y$)\cite{Christensen2018}.

	\subsubsection{Intercept}
		\label{rem:intercept}
		Note, that in contrast to ordinary linear regression, the model from \cref{eq:nd_eps} deliberately does \textit{not} include an intercept term, as a model with $S=\alpha + \textbf{x}^\top \boldsymbol{\beta} + \eps$ and shifted cut points $\tilde\theta_\ell = \theta_\ell-\alpha$ would result in the same distribution for the response $Y$. Thus, the cut-points $\theta_\ell$ would not be identifiable in a model including a fixed \textit{unknown} intercept. A model including a fixed \textit{known} intercept, however, would result in the same parameter estimates $\boldsymbol{\beta}$ as for the model (\ref{eq:nd_eps}). For this reason, without loss of generality $\alpha=0$ is assumed.

  \subsection{Interpretation of model parameters}
		\label{rem:interpretation}
		To conclude this section, let us briefly address the influence of the regression coefficients. An interpretation of this influence on the response variable $Y$ is, in general, difficult and depends on the chosen link function, cf.~Agresti\cite[Section 5.1.3]{Agresti2012}. For the logit-link, there is the following relation for the difference in log-odds (i.e.,\ the logarithm of the odds ratio). Note that here, the odds are the ratio of responding at most~$\ell$ vs.\ responding at least $\ell +1$, $\ell=1,2,\dots, L$. Denote the vector of covariates by $\textbf{x}=(x_1,x_2,\dots, x_K)$ and let $\tilde{\textbf{x}}=(x_1,\dots,x_{k-1}, x_k+1, x_{k+1}, \dots,x_K)$ be differing from $\textbf{x}$ only at the $k$-th component by one, then 
		\begin{align*}
			\log\left(\frac{\P(Y\le \ell \mid \textbf{x})}{1-\P(Y\le \ell \mid \textbf{x})}\right) - \log\left(\frac{\P(Y\le \ell \mid \tilde{\textbf{x}})}{1-\P(Y\le \ell \mid \tilde{\textbf{x}})}\right) &\stackrel{(\ref{eq:logodds})}{=} \theta_\ell - \textbf{x}^\top \boldsymbol{\beta} - \left( \theta_\ell - \tilde{\textbf{x}}^\top \boldsymbol{\beta} \right)\\
			&= \left(\tilde{\textbf{x}}-\textbf{x}\right)^\top \boldsymbol{\beta} = \beta_k,
		\end{align*} 
		for all $\ell=1,2,\dots,L$.

		Thus, for a binary group variable, the coefficient $\beta_k$ is the change in the log-odds for a change in groups for each cumulative probability, keeping all other parameters fixed. More general, for a continuous covariate $x_k$, $\beta_k$ is the effect of a \textit{unit increase} in $x_k$ on the log-odds for each cumulative probability, controlling for all the other predictors. As a result, $\exp(\beta_k)$ is a cumulative odds ratio using any collapsing of the ordinal response for values of $x_k$ that differ by 1 unit. The corresponding consequences for the question analysed and an interpretation beyond that are, however, often debatable. Here, the interpretation of predicted response probabilities is more recommended and represents the true strength of ordinal models.
		
		For an arbitrary distribution of $S$, we can still claim that a unit increase in $x_k$ corresponds to an increase in the mean $\E S$ of $S$ by $\beta_k$, keeping the other predictor values fixed, cf.\ Agresti\cite[Sections 5.1.3 and 5.2.1]{Agresti2012}. Stating a relation between the regression parameters and the response distribution in general is, however, not possible.

	\section{Identification of influencing factors and effect quantification}
	\label{sec:discrete_covariates}
	
\begin{figure}[h!]
		\includegraphics[width=\linewidth]{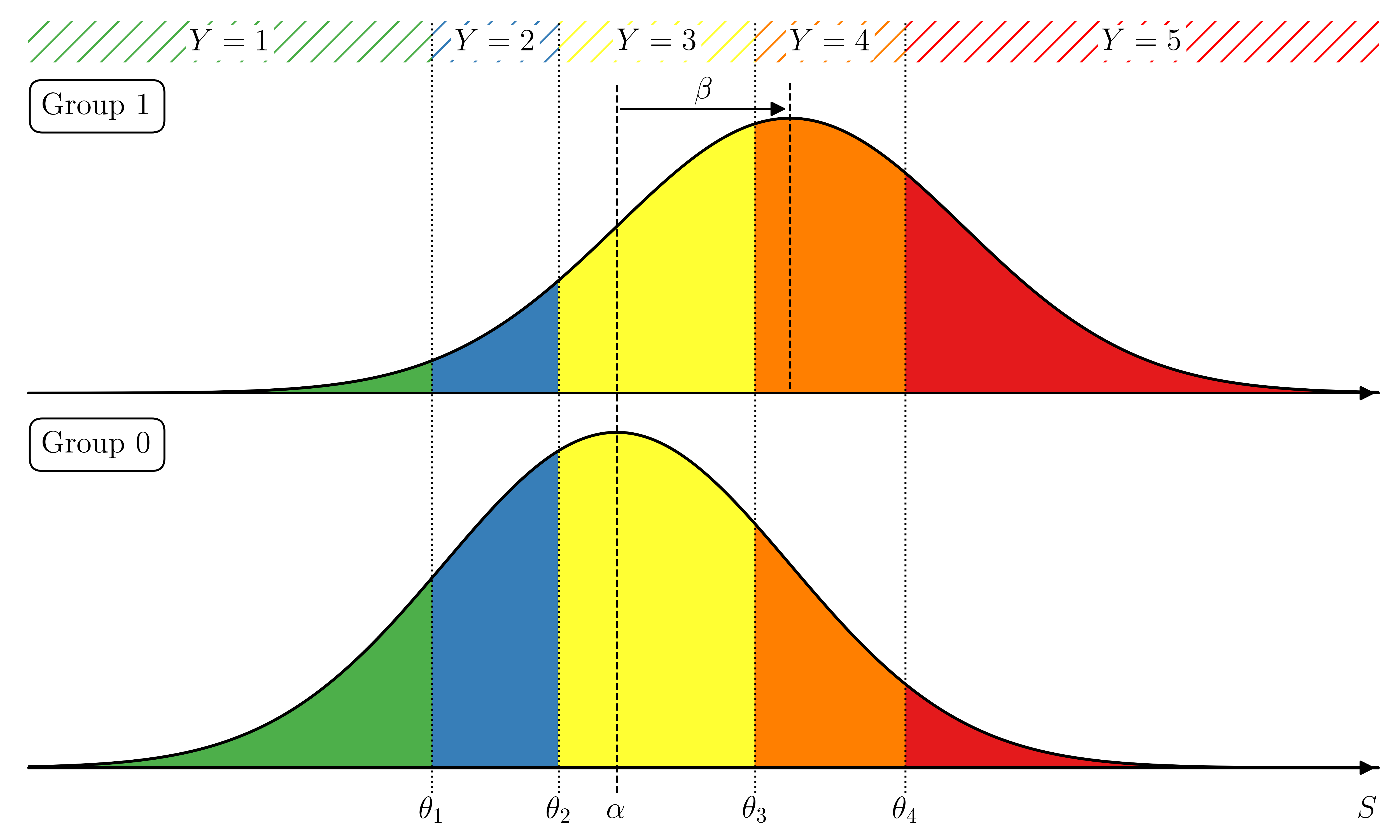}
		\caption{Graphical representation of the influence of a binary group variable to the distribution of the latent variable $S$ and effects on the probabilities of the survey responses~$Y$: The entire distribution of the latent variable, particularly its mean $\alpha$ (dashed lines), is shifted by the regression coefficient $\beta$ to the right (left) if $\beta >0$ ($\beta<0$). The probabilities (coloured areas) for all response levels change accordingly.}
		\label{fig:model_parameters}
	\end{figure}

	Let us now investigate the influences of covariates on the response outcome. To this end, we consider at first discrete covariates, say, a binary group variable having two levels 0/1 (e.g., two different studies with one question each having the same possible answers). Then, the distribution of the latent variable $S$ for group 1 is the one for group 0 shifted by the regression coefficient $\beta$, see \cref{fig:model_parameters}. This results in a shift of the probability masses for all possible responses. In the example, this means that positive values for $\beta$ make responses encoded with high values more likely (and responses encoded with low values less likely), and vice versa for negative values of $\beta$. 
	
	Using the framework presented above, we are moreover able to test for significant differences in the rating behaviour between two groups. Two rating behaviours are considered to differ significantly if the corresponding regression coefficient $\beta$ on the latent scale differs significantly from zero, i.e., there is a significant shift in the distribution of the latent variable~$S$. Testing procedures are of asymptotic Wald-type, cf.~Christensen\cite[Section 4.1]{Christensen2015}.
	
	\subsection{Model application}
		\label{ex:clm1}
		We consider the data from the \textsc{Filippa} study from \cref{sec:motivation}. We are interested whether and how the invasiveness of the study (observational/ RCT/ pharmacological study) and the child's sex influence the response behaviour of the legal representatives. Therefore, we fit an ordinal regression model for the response with covariates \texttt{study} type and child's \texttt{sex} with logit-link using the \texttt{clm} function from the \texttt{ordinal} package:

\begin{knitrout}
\definecolor{shadecolor}{rgb}{0.969, 0.969, 0.969}\color{fgcolor}\begin{kframe}
\begin{alltt}
\hlstd{clmodel} \hlkwb{<-} \hlkwd{clm}\hlstd{(response} \hlopt{~} \hlstd{study} \hlopt{+} \hlstd{sex,} \hlkwc{data} \hlstd{= data,} \hlkwc{link} \hlstd{=} \hlstr{'logit'}\hlstd{)}
\hlkwd{summary}\hlstd{(clmodel)}
\end{alltt}
\begin{verbatim}
## formula: response ~ study + sex
## data:    data
## 
##  link  threshold nobs logLik  AIC    niter max.grad cond.H 
##  logit flexible  318  -359.80 733.60 6(1)  5.47e-12 2.3e+02
## 
## Coefficients:
##             Estimate Std. Error z value Pr(>|z|)    
## studyRCT      2.2127     0.3723   5.943 2.79e-09 ***
## studyPharma   2.8428     0.3682   7.721 1.16e-14 ***
## sexMale      -0.4778     0.2350  -2.033    0.042 *  
## ---
## Signif. codes:  0 '***' 0.001 '**' 0.01 '*' 0.05 '.' 0.1 ' ' 1
## 
## Threshold coefficients:
##                                   Estimate Std. Error z value
## Absolutely consent|Rather consent   1.9020     0.3362   5.658
## Rather consent|Unsure               2.3924     0.3446   6.942
## Unsure|Rather decline               2.9745     0.3541   8.401
## Rather decline|Absolutely decline   3.6105     0.3670   9.837
\end{verbatim}
\end{kframe}
\end{knitrout}
		
		This model provides:
		\begin{itemize}
			\item $L-1 = 4$ threshold coefficients (cut-points), corresponding to the $\theta_{\ell}$'s in \cref{fig:model_parameters};
			\item two coefficients for the levels ``RCT'' (\texttt{studyRCT}) and ``pharmacological study'' \linebreak(\texttt{studyPharma}) of invasiveness (in comparison to the ``observational study'');
			\item and one coefficient \texttt{sexMale} for the influence of the child's sex (here: boys in comparison to girls).
		\end{itemize}
		The estimated coefficients for study invasiveness and sex are statistically significant (according to a Wald test, $p<0.001$ and $p=0.042$, respectively) at the two-sided 5\% significance level. As the coefficient for study type ``RCT'' is positive, the latent score distribution of $Y$ for an RCT study is shifted to the right by $2.2127$ in comparison to the observational study. See \cref{fig:model_parameters} for an illustration where 1 denotes the highest level of consent (``absolutely consent'') and 5 the least (``absolutely decline''). As the cut-points do not depend on the covariates, by this shift we predict levels with lesser consent with higher probability. 
	
		A similar argument holds for the pharmacological study compared to the observational study (here, the shift is $2.8428$, i.e.\ another $0.6301$ units w.r.t.\ ``RCT''). Moreover, as the coefficient for \texttt{sex} is negative ($-0.4778$), legal representatives of boys were significantly more likely to consent to participate in the studies than those of girls ($p=0.042$). 
		
		To assess whether there is also a significantly different response behaviour between the RCT and the pharmacological study, we have to conduct one additional test. Note that to this, we have to adjust for multiple testing. For this purpose, the \texttt{emmeans} method from the package of the same name \texttt{emmeans} \cite{Lenth2022} provides several possibilities using the \texttt{adjust} parameter. By default, $p$-value adjustment using Tukey's method \cite{Tukey1953}, for multiple comparisons is applied:		
\begin{knitrout}
\definecolor{shadecolor}{rgb}{0.969, 0.969, 0.969}\color{fgcolor}\begin{kframe}
\begin{alltt}
\hlkwd{emmeans}\hlstd{(clmodel,} \hlkwc{specs} \hlstd{=} \hlkwd{list}\hlstd{(pairwise} \hlopt{~} \hlstd{study),} \hlkwc{mode} \hlstd{=} \hlstr{'latent'}\hlstd{)[[}\hlnum{2}\hlstd{]]}
\end{alltt}
\begin{verbatim}
##  1                      estimate    SE  df z.ratio p.value
##  Observational - RCT       -2.21 0.372 Inf  -5.943  <.0001
##  Observational - Pharma    -2.84 0.368 Inf  -7.721  <.0001
##  RCT - Pharma              -0.63 0.252 Inf  -2.496  0.0336
## 
## Results are averaged over the levels of: sex 
## P value adjustment: tukey method for comparing a family of 3 estimates
\end{verbatim}
\end{kframe}
\end{knitrout}
		See also \cite[Section 5]{Hsu1996} for the theoretical background and a comprehensive overview about multiple testing problems. From the last column it follows that there are significant differences in the response behaviour of voters between all three study types at the significance level 5\%.
		
		Note that in the model summary there are no $p$-values provided for the threshold coefficients as testing against zero would not make much sense---the actual position of the cut-points has no meaning but only the distances between each other and the relative position to the mean of the latent variable, cf.~\cref{rem:intercept}.
		
		For this study, the odds for a legal representative of a girl to respond ``absolutely consent'' for the observational study are about $\exp(2.2127)=9.1$ times the odds for the RCT. This estimate is, however, relatively imprecise which can probably be reduced to the small number of legal representatives of a girl absolutely willing to participate in the RCT or the small sample size of the study overall. The 95\%-confidence interval is $[1.5154, 2.9858]$, corresponding to $[4.551, 19.803]$ for the odds ratio. As descriptive statistics of effects (here: probabilities) can compare the cumulative probabilities more suitably and is easier to interpret, we recommend to describe effects for ordinal regression quantitatively by presenting comparisons of probabilities e.g.\ at their extreme values, see also Agresti\cite[Section 8.2.4]{Agresti2012}. The procedure how to do so is described as follows.

	\subsection{Predicting probabilities, confidence intervals}
		The package \texttt{emmeans} provides moreover a method for computing predicted marginal response probabilities for each possible answer (columns \texttt{response} and \texttt{prob}), possibly stratified by given covariates (first line in each paragraph below):			
\begin{knitrout}
\definecolor{shadecolor}{rgb}{0.969, 0.969, 0.969}\color{fgcolor}\begin{kframe}
\begin{alltt}
\hlkwd{emmeans}\hlstd{(clmodel,} \hlopt{~} \hlstd{response} \hlopt{|} \hlstd{study} \hlopt{/} \hlstd{sex,} \hlkwc{mode} \hlstd{=} \hlstr{'prob'}\hlstd{)}
\end{alltt}
\begin{verbatim}
## study = Observational, sex = Female:
##  response             prob      SE  df asymp.LCL asymp.UCL
##  Absolutely consent 0.8701 0.03799 Inf   0.79565    0.9446
##  Rather consent     0.0461 0.01408 Inf   0.01853    0.0737
##  Unsure             0.0352 0.01162 Inf   0.01239    0.0579
##  Rather decline     0.0223 0.00799 Inf   0.00660    0.0379
##  Absolutely decline 0.0263 0.00941 Inf   0.00789    0.0448
...
\end{verbatim}
\end{kframe}
\end{knitrout}
		Note, that as the regression parameters and cut-points are subject to uncertainties (the data is random), so are the \textit{estimated probabilities} for the possible responses. Corresponding asymptotic (indicated by the infinite number of degrees of freedom \texttt{df=Inf}) confidence intervals, given as $[$\texttt{asymp.LCL}, \texttt{asymp.UCL}$]$, can be derived using the delta method and are provided in addition to the predictions in the \texttt{emmeans} method above (last two columns). We refer to Christensen\cite[Section 4.7]{Christensen2018} for further details on the derivation of standard errors (column \texttt{SE}) to compute these confidence intervals.

		The \texttt{emmeans} package contains moreover a function to show the regression output graphically using \texttt{emmip} for a clear and compact presentation:			
\begin{knitrout}
\definecolor{shadecolor}{rgb}{0.969, 0.969, 0.969}\color{fgcolor}\begin{kframe}
\begin{alltt}
\hlkwd{emmip}\hlstd{(clmodel,} \hlopt{~} \hlstd{response} \hlopt{|} \hlstd{study} \hlopt{/} \hlstd{sex,} \hlkwc{mode} \hlstd{=} \hlstr{'prob'}\hlstd{,} \hlkwc{CI} \hlstd{=} \hlnum{TRUE}\hlstd{,}
        \hlkwc{style} \hlstd{=} \hlstr{'factor'}\hlstd{,}
        \hlkwc{CIarg} \hlstd{=} \hlkwd{list}\hlstd{(}\hlkwc{colour}\hlstd{=response_colours,} \hlkwc{size}\hlstd{=}\hlnum{5}\hlstd{,} \hlkwc{alpha}\hlstd{=}\hlnum{.5} \hlstd{))} \hlopt{+}
  \hlkwd{labs}\hlstd{(}\hlkwc{y} \hlstd{=} \hlstr{'Estimated probability'}\hlstd{,} \hlkwc{x} \hlstd{=} \hlstr{'Level of consent'}\hlstd{)} \hlopt{+}
  \hlkwd{scale_x_discrete}\hlstd{()} \hlopt{+}
  \hlkwd{scale_y_continuous}\hlstd{(}\hlkwc{labels} \hlstd{= scales}\hlopt{::}\hlstd{percent)} \hlopt{+}
  \hlkwd{facet_grid}\hlstd{(sex} \hlopt{~} \hlstd{study)} \hlopt{+} \hlstd{theme}
\end{alltt}
\end{kframe}\begin{figure}[h!]
\includegraphics[width=\maxwidth]{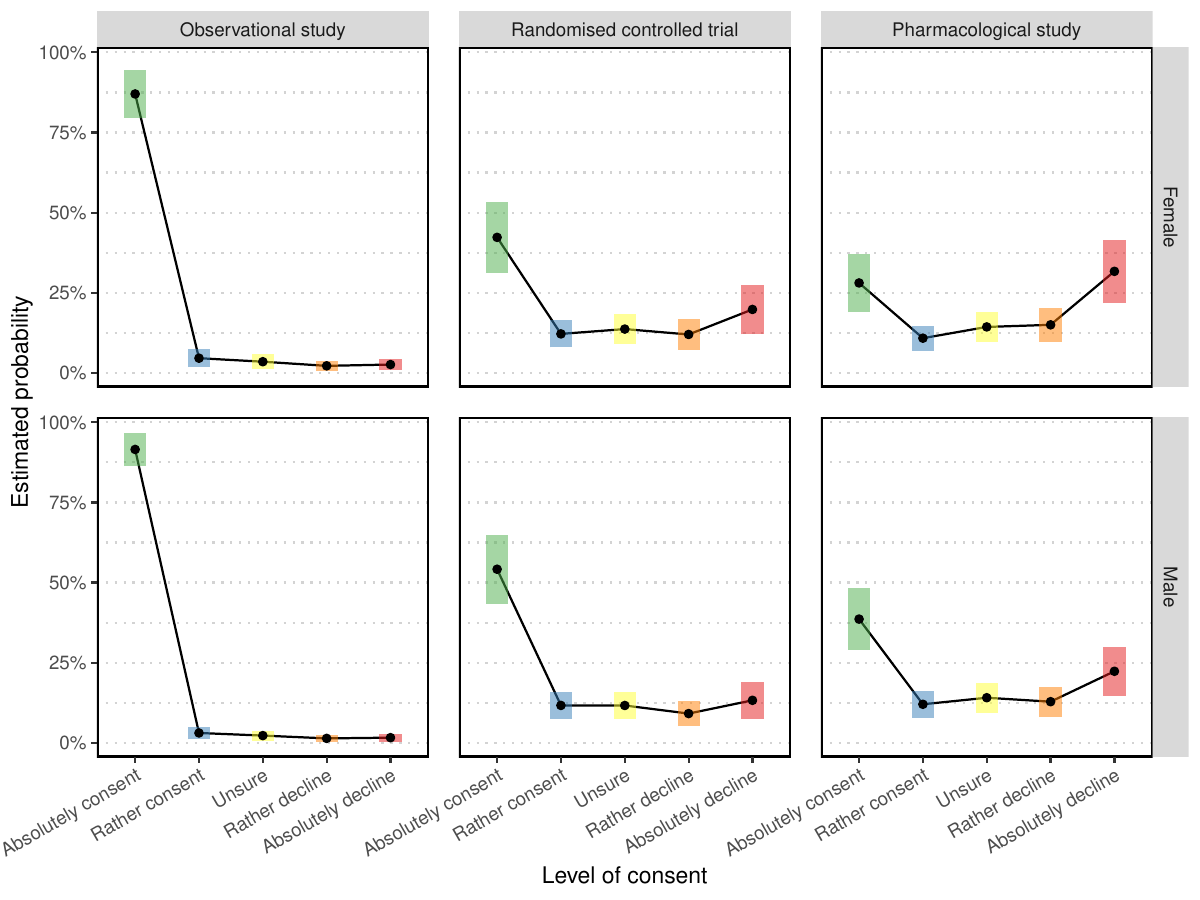} \caption{\label{fig:estimated_probs}Estimated response probabilities stratified by study type (horizontally) and child's sex (vertically) within the ordinal regression model of \cref{ex:clm1}.}\label{fig:unnamed-chunk-9}
\end{figure}

\end{knitrout}
		
		\cref{fig:estimated_probs} clearly shows differences in the response behaviour of legal representatives depending on the study invasiveness. More precisely, the statement that willingness to consent for participation decreases with increasing invasiveness from \cref{ex:clm1} is confirmed. Further, we discern smaller differences between the willingness to consent between boys and girls. Particularly for the RCT and the pharmacological study, more participants were responding ``absolutely consent'' and less ``absolutely decline'' in the boys' strata than in girls'. Differences were about 13\% and 6\% for the RCT and about 11\% and 10\% for the pharmacological study, respectively.
		
		Finally we like to remark that \texttt{emmeans} provides for the possibility to compute cumulative and exceedance (i.e.\ 1 $-$ cumulative) probabilities using \texttt{mode = 'cum.prob'} and \texttt{mode = 'exc.prob'}, respectively. For instance, the probability for a legal representative of at least "rather consenting" is:

\begin{knitrout}
\definecolor{shadecolor}{rgb}{0.969, 0.969, 0.969}\color{fgcolor}\begin{kframe}
\begin{alltt}
\hlkwd{emmeans}\hlstd{(clmodel,} \hlopt{~} \hlstd{study} \hlopt{/} \hlstd{sex,} \hlkwc{mode} \hlstd{=} \hlstr{'cum.prob'}\hlstd{,}
                                \hlkwc{at} \hlstd{=} \hlkwd{list}\hlstd{(}\hlkwc{cut} \hlstd{=} \hlstr{"Rather consent|Unsure"}\hlstd{))}
\end{alltt}
\begin{verbatim}
##  study                       sex    cumprob     SE  df asymp.LCL   asymp.UCL
##  Observational study         Female   0.916 0.0264 Inf     0.864   0.968
##  Randomised controlled trial Female   0.545 0.0576 Inf     0.432   0.658
##  Pharmacological study       Female   0.389 0.0533 Inf     0.285   0.494
##  Observational study         Male     0.946 0.0180 Inf     0.911   0.982
##  Randomised controlled trial Male     0.659 0.0504 Inf     0.560   0.757
##  Pharmacological study       Male     0.507 0.0510 Inf     0.407   0.607
##
## Confidence level used: 0.95
\end{verbatim}
\end{kframe}
\end{knitrout}

Note, that the cumulative and exceedance probablities can be obtained from the individual probability estimates above. This, however, does not hold for the confidence intervals.

For the model parameters, the \texttt{ordinal} package provides moreover profile likelihood based confidence intervals for model covariates. These confidence intervals often possess better coverage than Wald-type confidence intervals, particularly in studies with small to moderate sample size. The computation can be performed via:

\begin{knitrout}
\definecolor{shadecolor}{rgb}{0.969, 0.969, 0.969}\color{fgcolor}\begin{kframe}
\begin{alltt}
\hlkwd{confint}\hlstd{(clmodel)}
\end{alltt}
\begin{verbatim}
##                                      2.5 %      97.5 %
## studyRandomised controlled trial  1.515385  2.98583513
## studyPharmacological study        2.156130  3.60999306
## sexMale                          -0.940631 -0.01824491
\end{verbatim}
\end{kframe}
\end{knitrout}
	
	Profile likelihood based confidence intervals are implemented for regression and scale parameters, but are not available for threshold, nominal and flexible link parameters \cite{Christensen2018}.
		
	\subsection{Numerics and convergence}
		\label{rem:numerics}
		The model coefficients are estimated numerically using a maximum likelihood approach by calculation of zeroes of the gradient of the negative log-likelihood. Parameter estimates are output after a convergence criterion is satisfied (typically small gradient) or a maximum number of iterations has been reached. Numerics and control parameters can be passed using \texttt{clm.control}.
		
		To oversee the convergence of the approach, the \texttt{summary} of the \texttt{clm} method provides three parameters, see \cref{ex:clm1}: \texttt{niter} (the number of Newton-Raphson iterations needed with the number of step-halvings in parentheses), \texttt{max.grad} (the maximum absolute gradient of log-likelihood) and \texttt{cond.H} (the condition of the Hessian at the maximum). More detailed information can be obtained using the \texttt{convergence} method applied to the model (cf.\ also \cref{rem:convergence} below). Christensen\cite{Christensen2015} states that large \texttt{cond.H} values (like $>$ \texttt{1e4}) might indicate that the model is ill-defined. As in the case of \cref{ex:clm1} \texttt{max.grad} is small and \texttt{cond.H} is reasonably sized, we can conclude that the algorithm seems to have converged properly.
	
	\subsection{Interaction effects}
		\label{rem:interaction}
		One might wonder, whether it is appropriate to include also an interaction term \texttt{study:sex} in the model. To this end, we can fit a model including this interaction and compare it with the model from \cref{ex:clm1} (not containing factor interaction). For this purpose, the \texttt{ordinal} package provides the \texttt{anova} method conducting a likelihood ratio test between both models:
\begin{knitrout}
\definecolor{shadecolor}{rgb}{0.969, 0.969, 0.969}\color{fgcolor}\begin{kframe}
\begin{alltt}
\hlstd{clmodel_inter} \hlkwb{<-} \hlkwd{clm}\hlstd{(response} \hlopt{~} \hlstd{study} \hlopt{*} \hlstd{sex,} \hlkwc{data} \hlstd{= data,} \hlkwc{link} \hlstd{=} \hlstr{'logit'}\hlstd{)}
\hlkwd{anova}\hlstd{(clmodel, clmodel_inter)}
\end{alltt}
\begin{verbatim}
## Likelihood ratio tests of cumulative link models:
##  
##               formula:               link: threshold:
## clmodel       response ~ study + sex logit flexible  
## clmodel_inter response ~ study * sex logit flexible  
## 
##               no.par    AIC  logLik LR.stat df Pr(>Chisq)
## clmodel            7 733.60 -359.80                      
## clmodel_inter      9 737.29 -359.65   0.307  2     0.8577
\end{verbatim}
\end{kframe}
\end{knitrout}
		This implies that there is no statistical evidence for including interaction \texttt{study:sex} between study-type and child's sex into the model ($p=0.8577$).

	\subsection{Random effects}
		\label{rem:random}
		To conclude this first example, we consider the question whether there are substantial differences in the individual response behaviours, i.e.\ whether there are parents responding systematically e.g. particularly low or high values or answering always ``unsure" etc. To this end, we fit an ordinal model as in \cref{ex:clm1} but with additional individual random effect using the \texttt{clmm} function from the \texttt{ordinal} package and compare this model, analogously to \cref{rem:interaction} with the model from \cref{ex:clm1}:
\begin{knitrout}
\definecolor{shadecolor}{rgb}{0.969, 0.969, 0.969}\color{fgcolor}\begin{kframe}
\begin{alltt}
\hlstd{clmmodel} \hlkwb{<-} \hlkwd{clmm}\hlstd{(response} \hlopt{~} \hlstd{study} \hlopt{+} \hlstd{sex} \hlopt{+} \hlstd{(}\hlnum{1} \hlopt{|} \hlstd{id),} \hlkwc{data} \hlstd{= data,}
                            \hlkwc{link} \hlstd{=} \hlstr{'logit'}\hlstd{)}
\hlkwd{anova}\hlstd{(clmodel, clmmodel)}
\end{alltt}
\begin{verbatim}
## Likelihood ratio tests of cumulative link models:
##  
##          formula:                          link: threshold:
## clmodel  response ~ study + sex            logit flexible  
## clmmodel response ~ study + sex + (1 | id) logit flexible  
## 
##          no.par    AIC  logLik LR.stat df Pr(>Chisq)
## clmodel       7 733.60 -359.80                      
## clmmodel      8 733.47 -358.74  2.1299  1     0.1444
\end{verbatim}
\end{kframe}
\end{knitrout}
		Again, there is no strong evidence ($p=0.1444$) for substantial subject specific response behaviours; a test at the 5\% significance level would not reject the model from \cref{ex:clm1} in favour to a model including additional individual random effects.
		
		Note that, as of now, the \texttt{clmm} function does not support predicting probabilities of a model containing random effects. In contrast, the former implementation \texttt{clmm2} of \texttt{clmm} does support prediction, however, provides only fitted values for a random effect of zero\cite{Christensen2015}. Beyond this, we present an approximate approach using the \texttt{emmeans} method at the end of the following \cref{ex:clm2}.
		
		Of note, methods from the \texttt{ordinal} package seem to often not converge when there are random effects included. In this case, it might be worthwhile to consider e.g.\ a Bayesian random effects ordinal logistic model, cf.\ the \texttt{blrm} from the \texttt{rmsb} package \cite{HarrellJr2023b}. Moreover, a further alternative for including repeated measures into ordinal models is implemented in the \texttt{repolr} package \cite{Parsons2016} which extends the \texttt{polr} from the \texttt{MASS} package \cite{Venables2002}. More sophisticated designs for analysing survey data often involve clustering and probability sampling weights. We refer the interested reader to An et al.; Selosse et al.\cite{An2002, Selosse2021} or Agresti\cite[Chapter 12ff.]{Agresti2012}.

	\subsection{Continuous covariates}
		\label{ex:clm2}
\begin{figure}[h!]
			\includegraphics[width=\linewidth]{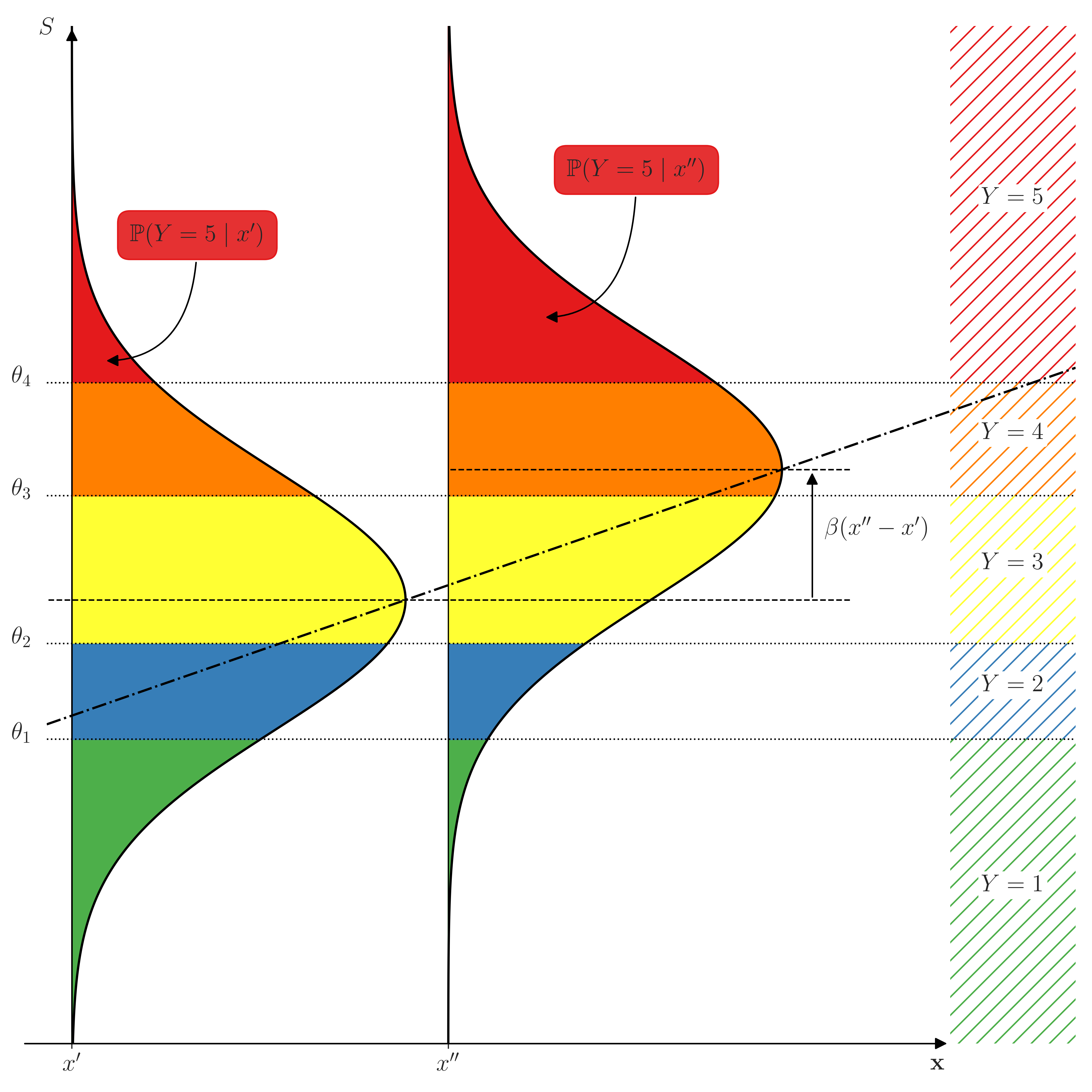}
			\caption{Graphical representation of the influence of a univariate continuous covariate $\textbf{x}\in\R^1$ to the distribution of the latent variable $S$ and effects on the probabilities of the survey responses~$Y$: A change of the covariate $\textbf{x}$ causes a change of the mean of the latent score $\E S$ (cf. $\alpha$ in \cref{fig:model_parameters}) along the regression line (dash-dotted $-\cdot$), i.e.\ if the line's slope is $\beta$, a change of $\textbf{x}$ by one unit results in a change of $\E S$ by $\beta$ units. Particularly, assuming fixed cut-points, if the regression coefficient $\beta$ is positive (negative), a higher value of $\textbf{x}$ results in tendentially higher (smaller) response values.}
			\label{fig:continuous_covariate}
		\end{figure}

		Considering the influence of a continuous covariate on the response outcome, fitting of models is done analogously as described in \cref{ex:clm1} above. The influence of a continuous covariate is portrayed schematically in \cref{fig:continuous_covariate} for a univariate covariate ${x}\in\R^1$. The covariate ${x}$ is allowed to vary along the horizontal axis whereas the latent score is plotted on the vertical axis. This corresponds to \cref{fig:model_parameters} reflected at the 45\textsuperscript{$\circ$}~line. A change in ${x}$ by one unit results in a shift of the latent score distribution by the corresponding regression coefficient, i.e.\ here~$\beta$ units. More generally, the mean of the latent score between two responders with covariates ${x}'$ and ${x}''$ is shifted by $\pm{\beta}({x}''-{x}')$ (depending on which direction one considers). The probabilities of the response categories shift accordingly, given unchanged cut-points, see the coloured areas in \cref{fig:continuous_covariate}.
		
		Below you can find the result for an ordinal model for the response depending on study type and child's sex, and additionally the age of the legal representative for the data from the \textsc{Filippa} study. Note that whenever there was more than one legal representative present at the interview, we chose the age of the oldest one for this analysis.

\begin{knitrout}
\definecolor{shadecolor}{rgb}{0.969, 0.969, 0.969}\color{fgcolor}\begin{kframe}
\begin{alltt}
\hlstd{clmodel} \hlkwb{<-} \hlkwd{clm}\hlstd{(response} \hlopt{~} \hlstd{study} \hlopt{+} \hlstd{sex} \hlopt{+} \hlstd{partner_age,} \hlkwc{data} \hlstd{= data,} \hlkwc{link} \hlstd{=} \hlstr{'logit'}\hlstd{)}
\hlkwd{summary}\hlstd{(clmodel)}
\end{alltt}
\begin{verbatim}
## formula: response ~ study + sex + partner_age
## data:    data
## 
##  link  threshold nobs logLik  AIC    niter max.grad cond.H 
##  logit flexible  306  -342.41 700.82 6(1)  6.29e-10 1.7e+05
## 
## Coefficients:
##             Estimate Std. Error z value Pr(>|z|)    
## studyRCT     2.28547    0.38863   5.881 4.08e-09 ***
## studyPharma  3.00952    0.38550   7.807 5.87e-15 ***
## sexMale     -0.52753    0.24277  -2.173   0.0298 *  
## partner_age -0.03713    0.01546  -2.402   0.0163 *  
## ---
## Signif. codes:  
## 0 '***' 0.001 '**' 0.01 '*' 0.05 '.' 0.1 ' ' 1
## 
## Threshold coefficients:
##                                   Estimate Std. Error    z value
## Absolutely consent|Rather consent   0.6098     0.6546    0.931
## Rather consent|Unsure               1.0961     0.6564    1.670
## Unsure|Rather decline               1.7115     0.6573    2.604
## Rather decline|Absolutely decline   2.3568     0.6602    3.570
## (12 observations deleted due to missingness)
\end{verbatim}
\end{kframe}
\end{knitrout}
		
		We observe that there are only minor changes in the coefficients \texttt{studyRCT}, \texttt{studyPharma} and \texttt{sex} in comparison to \cref{ex:clm1}. The coefficient \texttt{partner\char`_age} states that for every increase in the age of the legal representative by one year, the mean latent score decreases by about $-0.03713$ units. This means that the older the interview partners are, the more likely they are responding smaller response levels, i.e.\ the more likely they are consenting to study participation.

		The resulting matrix of estimates can then be visualised as a stream plot as follows:

\begin{knitrout}
\definecolor{shadecolor}{rgb}{0.969, 0.969, 0.969}\color{fgcolor}\begin{kframe}
\begin{alltt}
\hlkwd{ggplot}\hlstd{(clm_fit,} \hlkwd{aes}\hlstd{(}\hlkwc{x} \hlstd{= partner_age,} \hlkwc{y} \hlstd{= fit,} \hlkwc{fill} \hlstd{= response))} \hlopt{+}
  \hlkwd{geom_stream}\hlstd{(}\hlkwc{type} \hlstd{=} \hlstr{'proportional'}\hlstd{,} \hlkwc{alpha} \hlstd{=} \hlnum{0.8}\hlstd{)} \hlopt{+}
  \hlkwd{labs}\hlstd{(}\hlkwc{y} \hlstd{=} \hlstr{'Estimated probability'}\hlstd{,} \hlkwc{x} \hlstd{=} \hlstr{'Age of legal representative [years]'}\hlstd{,}
  \hlkwc{fill} \hlstd{=} \hlstr{'Level of consent'}\hlstd{)} \hlopt{+}
  \hlkwd{scale_x_continuous}\hlstd{(}\hlkwc{expand} \hlstd{=} \hlkwd{c}\hlstd{(}\hlnum{0}\hlstd{,}\hlnum{0}\hlstd{),} \hlkwc{limits} \hlstd{=} \hlkwd{c}\hlstd{(}\hlnum{20}\hlstd{,} \hlnum{60}\hlstd{))} \hlopt{+}
  \hlkwd{scale_y_continuous}\hlstd{(}\hlkwc{expand} \hlstd{=} \hlkwd{c}\hlstd{(}\hlnum{0}\hlstd{,}\hlnum{0}\hlstd{),} \hlkwc{labels} \hlstd{= scales}\hlopt{::}\hlkwd{percent_format}\hlstd{())} \hlopt{+}
  \hlkwd{scale_fill_manual}\hlstd{(}\hlkwc{values} \hlstd{= response_colours,} \hlkwc{labels} \hlstd{= responses)} \hlopt{+}
  \hlstd{theme} \hlopt{+}
  \hlkwd{facet_grid}\hlstd{( sex} \hlopt{~} \hlstd{study)}
\end{alltt}
\end{kframe}\begin{figure}[h!]
\includegraphics[width=\maxwidth]{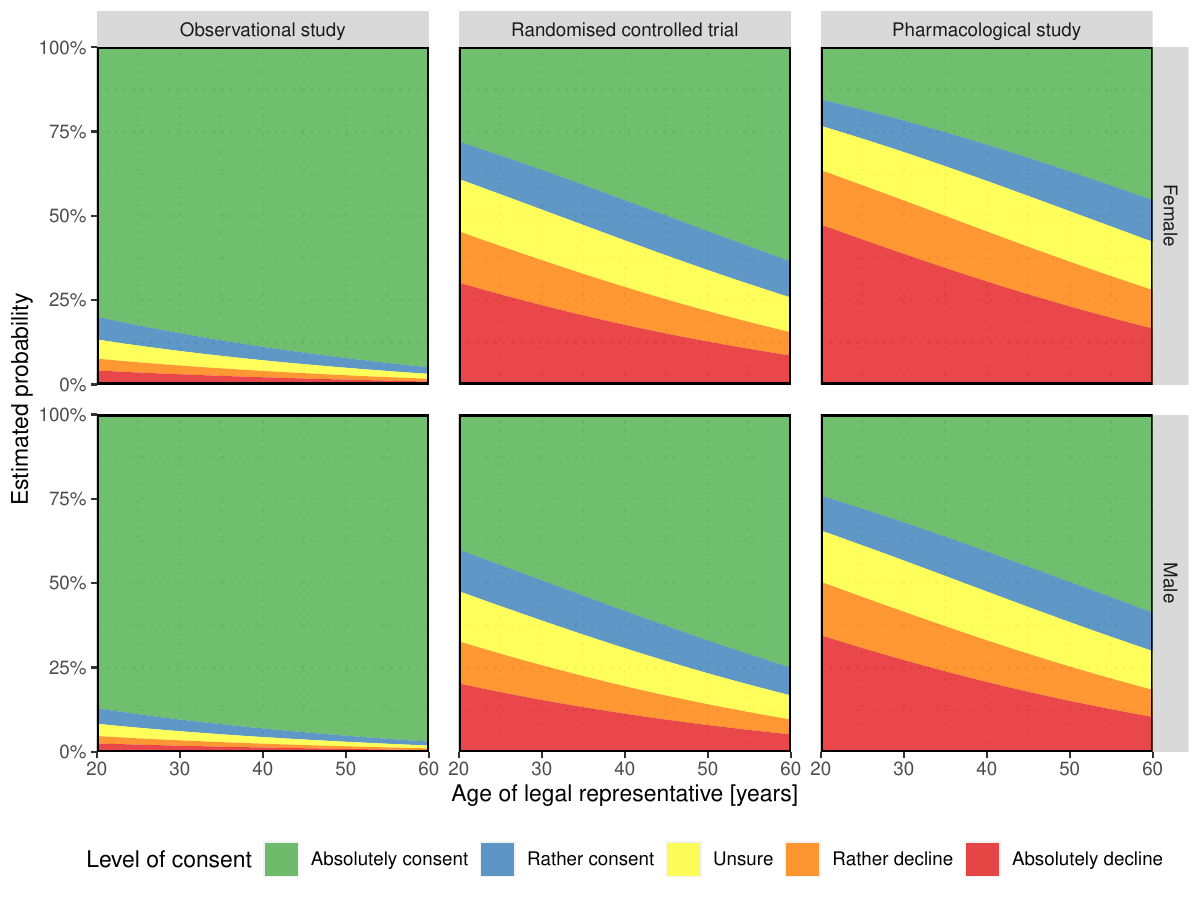} \caption{\label{fig:streamplot}Estimated response probabilities stratified by study type (horizontally) within the ordinal regression model of \cref{ex:clm2}.}\label{fig:unnamed-chunk-20}
\end{figure}

\end{knitrout}
		
		In all three studies the level of consent is increasing for an increasing age of the legal representative even though this is a bit more pronounced in male than in female inpatients, see~\cref{fig:streamplot}. If we included an interaction term into the regression model, e.g., \texttt{study:sex} between study type and child's sex, it might occur that this behaviour differs among the different study types. However, as neither interaction between \texttt{study}, \texttt{sex} and \texttt{partner\char`_age} was significant we did not include parameter interaction into the model.
		
		Note that \cref{fig:streamplot}, in contrast to \cref{fig:estimated_probs} does not contain any information about the estimated uncertainty of estimates (e.g., in form of confidence intervals). This is, however, rather a limitation of the graphical representation than of the applied method. We can still obtain asymptotic Wald-type confidence intervals for the probabilities of each response level given the covariates (e.g., here for all studies and a legal representative of age 42) using:
		
\begin{knitrout}
\definecolor{shadecolor}{rgb}{0.969, 0.969, 0.969}\color{fgcolor}\begin{kframe}
\begin{alltt}
\hlkwd{emmeans}\hlstd{(clmodel,} \hlopt{~} \hlstd{response} \hlopt{|} \hlstd{study} \hlopt{/} \hlstd{partner_age,} \hlkwc{mode} \hlstd{=} \hlstr{'prob'}\hlstd{,}
                \hlkwc{at} \hlstd{=} \hlkwd{list}\hlstd{(}\hlkwc{partner_age}\hlstd{=}\hlnum{42}\hlstd{))}
\end{alltt}
\begin{verbatim}
## study = Observational study, partner_age = 42:
##  response             prob      SE  df asymp.LCL asymp.UCL
##  Absolutely consent 0.9171 0.02658 Inf   0.86504    0.9692
##  Rather consent     0.0301 0.01016 Inf   0.01022    0.0500
##  Unsure             0.0235 0.00833 Inf   0.00716    0.0398
##  Rather decline     0.0137 0.00524 Inf   0.00341    0.0239
##  Absolutely decline 0.0156 0.00598 Inf   0.00386    0.0273
...
\end{verbatim}
\end{kframe}
\end{knitrout}
	
	\subsection{Invariance to choice of response categories}
		Agresti\cite[Section 3.3.3]{Agresti2012} points out that the regression parameters $\boldsymbol{\beta}$ in the latent variable model do \textit{not} depend on the particular way the continuous latent scale is cut by the cut-points $\theta_\ell$. Thus, the effect of the parameters $\boldsymbol{\beta}$ are independent of the choice of the categories of $Y$. For instance, the same effect parameters apply for a variable with five consent levels (as in \cref{ex:clm1} and \cref{ex:clm2} above) or to a response variable with e.g.\ ten or only three consent levels. This makes it possible to compare model parameters from studies using different scales.

	\section{Advanced topics}
	\label{sec:advanced_topics}
		
		To conclude we like to allude to a number of advanced topics which we deem to be important in the context fitting of ordinal models.
		
		\subsection{Convergence check}
			\label{rem:convergence}
			As already mentioned in \cref{rem:numerics}, the model coefficients are determined numerically. Whereas the \texttt{summary} command shows only a brief outline about model coefficients and convergence, the \texttt{convergence} command from the \texttt{ordinal} packages yields more comprehensive overview and provides additional information about the number of correctly estimated decimals. 
			
%			For demonstration, we exemplarily reexamine \cref{ex:clm2}. A glimpse in the convergence summary reveals that the model converged properly even though the condition of the Hessian of the log-likelihood is quite large.	
%
%			<<echo=TRUE, cache=TRUE, output.lines=-(1:10)>>=
%			convergence(clmodel)
%			@
%			
%			The change of the response distribution can be illustrated over time as follows: We predict for every combination of covariates on a discrete grid the corresponding response probability:

%			<<echo=TRUE, cache=TRUE, output.lines=7>>=
%			grid <- expand.grid(study = c("Observational", "RCT", "Pharma"), 
%			sex = c("Girls", "Boys"), partner_age = seq(20, 60, by = 1), 
%			response = responses)		
%			clm_fit <- data.frame(grid, fit = predict(clmodel, newdata = grid))
%			head(clm_fit)
%			@
		
		\subsection{Goodness of fit, model selection}
			In applications after model fitting typically the question for \textit{goodness of fit} arises, i.e.\ ``How well does our model predict the present data?''. For ordinary linear regression, quantities such as an $R^2$ coefficient of determination, quantifying the amount of data variance is explained by the model, are often stated. As a result, goodness of fit testing is enabled. In a nutshell, a non-significant $p$-value of a goodness of fit test indicates that there is no evidence that observed and fitted values/ frequencies do differ in a statistically significant way (thus indicating a reasonable fit). Comparisons of several nested models are possible using criteria such as the Akaike (AIC) or the Bayesian information criterion (BIC) to account for increasing goodness of fit for an increasing number of included parameters, cf.~Agresti\cite[Section 4.6]{Agresti2015}. 
			
			Goodness of fit testing in ordinal models is considered by Pulkstenis \& Robinson; Fagerland \& Hosmer\cite{Pulkstenis2004,Fagerland2016}. Corresponding algorithms are implemented in \texttt{R} in the \texttt{generalhoslem} package \cite{Jay2019}. For practical applications, Fagerland \& Hosmer\cite{Fagerland2013} recommend calculation of three different methods, the Lipsitz test, an ordinal version of the Hosmer-Lemeshow test, and the Pulkstenis-Robinson test, to assess goodness of fit, each covering slightly different aspects of the problem. Of note, these tests, however, do not have good power to detect a particular type of lack of fit \cite{Hosmer1997, Agresti2013}. Particularly, a large value of the global fit statistic only indicates some lack of fit, but does not provide insights about its nature \cite{Agresti2013}. For an elaborate discussion about these tests and further details we refer to Fagerland \& Hosmer\cite[Section 6]{Fagerland2013} or Hosmer Jr. et al.\cite[Chapter 5]{HosmerJr2013}.

Harrell Jr.\cite[Sections 13.3.5 and 13.3.6]{HarrellJr2015} propose checking model assumptions and model fit graphically, e.g. using nomograms for assessing the proportional odds assumption and quantifying the model's predictive ability using pseudo $R^2$ coefficients. Moreover, goodness-of-fit can be assessed by comparing the log-likelihood of the model with the one of a hoped-for-simpler model or the one of a richer model using tests similarly to the ones presented in \cref{rem:interaction} and \cref{rem:random}.

Finally, we like to remark that there is a number of various pseudo $R^2$ measures and measures of predictive discrimination for ordinal models. A common variant is Nagelkerke's\cite{Nagelkerke1991} $R^2$ which is also part of the standard output of the \texttt{lrm} function \cite{HarrellJr2023c}. A broader variety is provided by the \texttt{PseudoR2} function in the \texttt{DescTool} package \cite{Signorell2023}. We refer to Tjur \cite{Tjur2009} for a literature summary and some theoretical background.
			
		\subsection{Trial design and sample size computation}	
		
		In practice, often the question arises how many participants a study should include. In case of an ordinal analysis the \texttt{Hmisc} package \cite{HarrellJr2023a} provides the \texttt{posamsize} and \texttt{popower} functions to determine power of tests and sample size estimates for ordinal proportional odds models. E.g. for a comparison of the response behaviour between study types "Observational" and "RCT" at an expected response distribution of $p = (.87, .05, .04, .02 ,.02)^\top$ for the observational study and an expected odds ratio of 9 a two-sided test on the significance level 5\% achieves a power of 80\% if we include 58 patients:

\begin{knitrout}
\definecolor{shadecolor}{rgb}{0.969, 0.969, 0.969}\color{fgcolor}\begin{kframe}
\begin{alltt}
  \hlstd{Hmisc}\hlopt{::}\hlkwd{posamsize}\hlstd{(}\hlkwc{p} \hlstd{=} \hlkwd{c}\hlstd{(}\hlnum{.87}\hlstd{,} \hlnum{.05}\hlstd{,} \hlnum{.04}\hlstd{,} \hlnum{.02}\hlstd{,} \hlnum{.02}\hlstd{),} \hlkwc{odds.ratio} \hlstd{=} \hlnum{9}\hlstd{)}
\end{alltt}
\begin{verbatim}
## Total sample size: 57.2 
## Efficiency of design compared with continuous response: 0.341
\end{verbatim}
\end{kframe}
\end{knitrout}
			
		\subsection{Model generalisations}
			
\begin{figure}[H]
				\centering
				\begin{subfigure}{\linewidth}
					\centering
					\includegraphics[width=.9\linewidth]{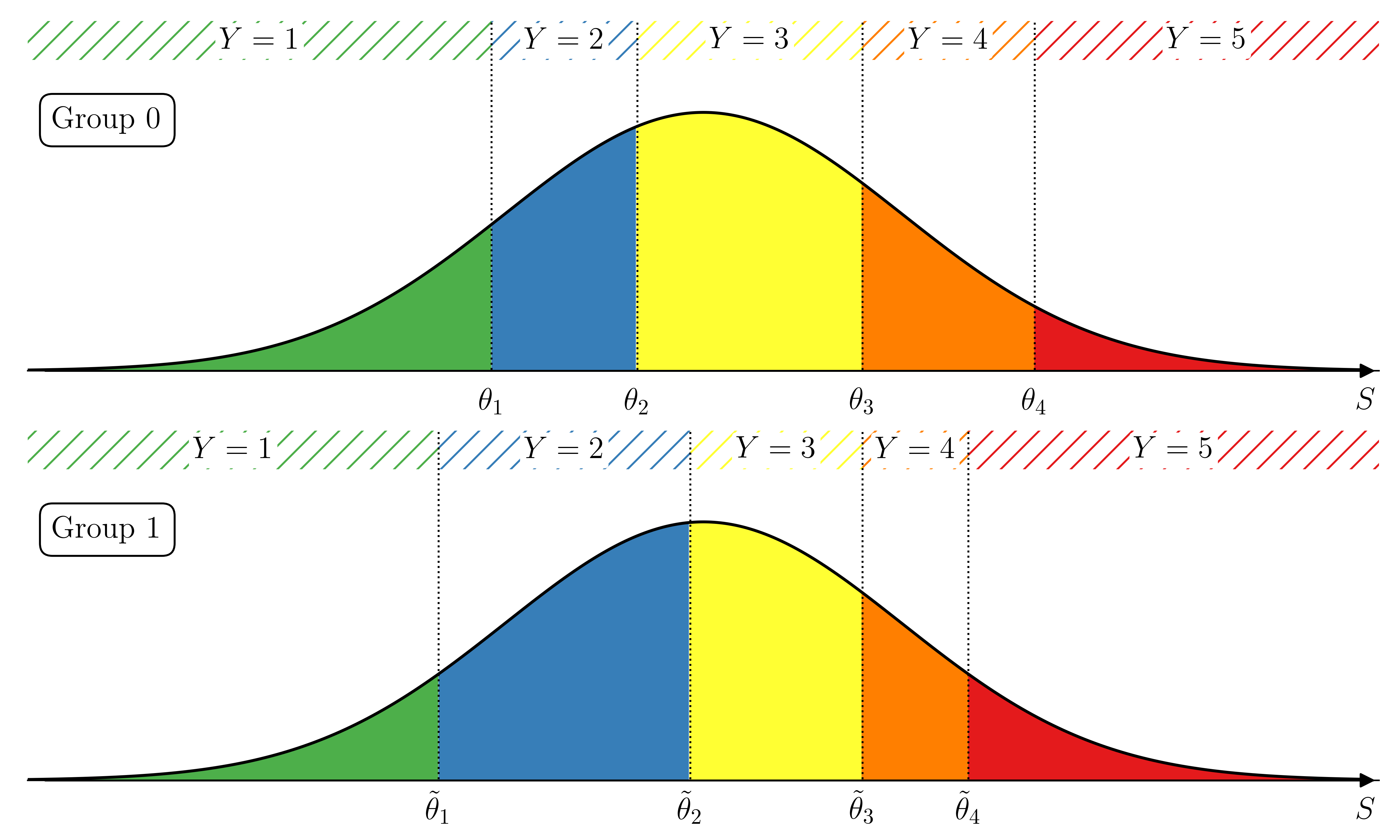}
					\caption{}
					\label{fig:nominal_effects}
				\end{subfigure}   
				\begin{subfigure}{\linewidth}
					\centering
					\includegraphics[width=.9\linewidth]{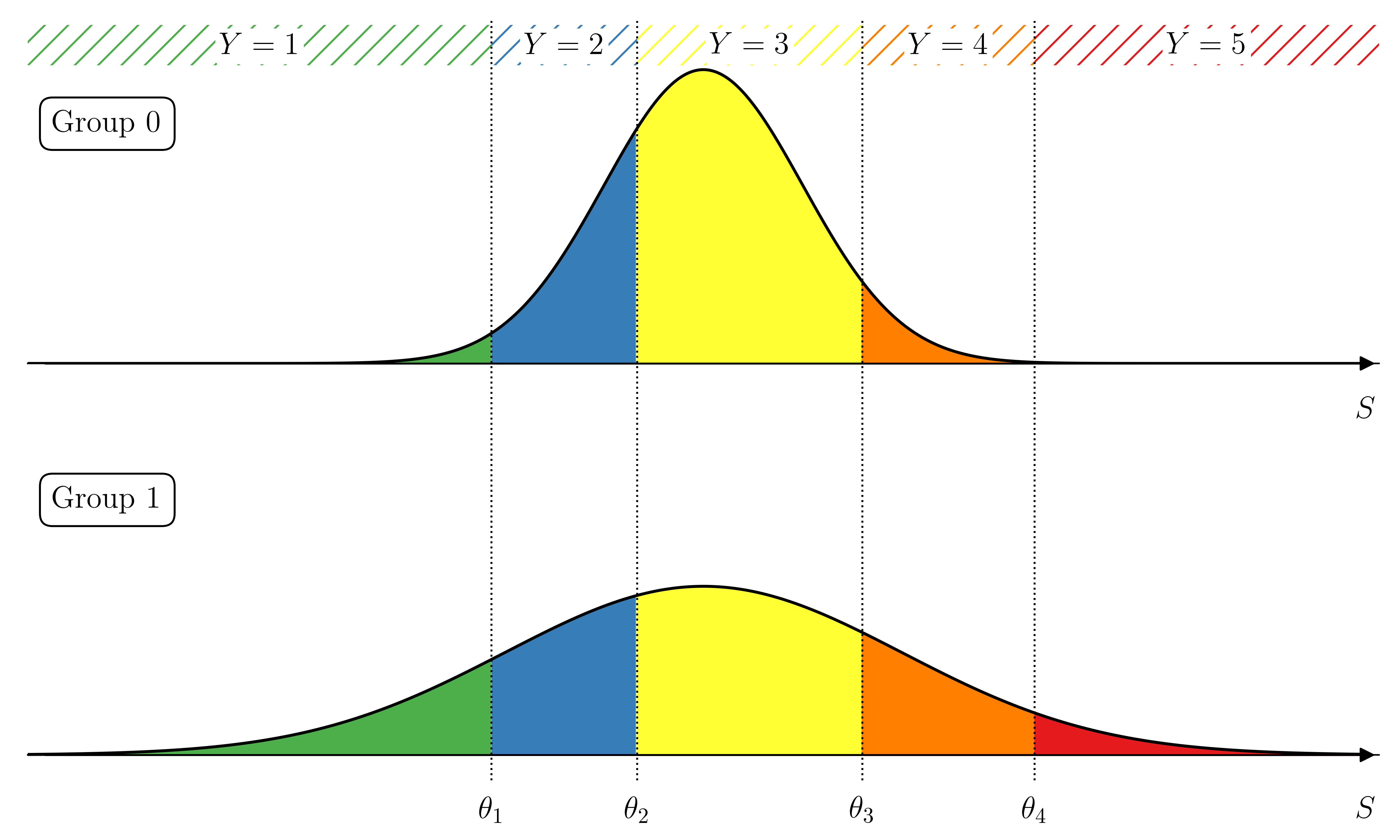}
					\caption{}
					\label{fig:scale_effects}
				\end{subfigure}
				\caption{Effects of including (a) nominal and (b) scale effects to an ordinal regression model. (a)~For every nominal effect we have to estimate another $L$ (number of response levels) regression parameters, or equivalently $L$ threshold coefficients $\tilde{\theta}_\ell = \theta_\ell - \tilde{\beta}_\ell$. (b)~Including scale effects changes the variance of the underlying latent variable $S$.}
				\label{fig:advanced_effects}
			\end{figure}

			Finally, the \texttt{ordinal} package provides a number of generalisations of the ordinal model presented in \cref{sec:ordinal_models}. The general form of a cumulative link model (see \cref{eq:clm}) can be written as
			\begin{align}
				\P(Y \le \ell \mid \textbf{x}, \textbf{w}, \textbf{z} ) &= F_\lambda\left( \frac{ g_{\boldsymbol{\alpha}}(\theta_\ell) - \textbf{x}^\top \boldsymbol{\beta} - \textbf{w}^\top \tilde{\boldsymbol{\beta}}_\ell}{\exp(\textbf{z}^\top \boldsymbol{\zeta})} \right) %= F_\lambda\left( \frac{ g_{\boldsymbol{\alpha}}(\theta_\ell) - \sum_{i=1}^k x_i \beta_i - \sum_{j=1}^m w_j \tilde{\beta}_{\ell, j}}{\exp\left(\sum_{s=1}^S z_s \zeta_s\right)} \right)
			\end{align}
			where the parameters affect the model as follows:
			\begin{itemize}
				\item $F_\lambda$ is the inverse link function which may be parametrised by a parameter $\lambda \in \R$. Its inverse $F_\lambda^{-1}$ is also referred to as \textit{flexible link function}.
				\item Cut-points may be transformed via $g_{\boldsymbol{\alpha}}(\theta_\ell)$ to be more \textit{structured} to reduce the number of model parameters, thus increasing efficiency of the estimators. Typical choices are assumptions of symmetric distribution of the cut-points of around the mean or having equal distances between each other (equidistant cut-points). Unrestricted cut-points (corresponding to $g_{\boldsymbol{\alpha}}$ to be the identity) are also referred to as \textit{flexible}.
				\item $\textbf{x}^\top \boldsymbol{\beta}$ are the ordinary regression effects as in \cref{eq:clm}.
				\item Regression effects (or the cut-points, respectively) might be allowed to depend on covariates to include \textit{nominal effects} $\textbf{w}^\top \tilde{\boldsymbol{\beta}}_\ell$, see \cref{fig:nominal_effects}. This allows for more flexibility in the modelling of rating behaviours. Computationally, to include nominal effects one has to pass the corresponding variable to the \texttt{nominal} parameter of \texttt{clm}. Note, however, that parameters included in nominal effects cannot be included as covariates due to identifiability reasons. Models including these nominal effects with logit-link are also called \textit{partial} or \textit{non-proportional odds} models \cite{Peterson1990}.
				\item The variance of the latent variable might be depending on covariates via $\exp(\textbf{z}^\top \boldsymbol{\zeta})$ (\textit{scale effects}, see \cref{fig:scale_effects}), e.g., reflecting different variances in the rating behaviour depending on group membership. To use scale effects in the \texttt{clm} method, pass variable names to the \texttt{scale} parameter while fitting the model.
			\end{itemize}
			
			A more comprehensive overview is provided in Christensen \cite[Section 2.3ff.]{Christensen2018} with corresponding remarks concerning implementations in Section~4\cite{Christensen2018}. Whether it is reasonable to include any of these modifications into the model is usually up to the analysing statistician. Models can again be compared using likelihood ratio tests as described in \cref{rem:interaction} and \cref{rem:random}, respectively. For a fitted model, likelihood ratio tests for models after adding nominal or scale effects can be performed via

\begin{knitrout}
\definecolor{shadecolor}{rgb}{0.969, 0.969, 0.969}\color{fgcolor}\begin{kframe}
\begin{alltt}
  \hlkwd{nominal_test}\hlstd{(clmodel)}
\end{alltt}
\begin{verbatim}
## Tests of nominal effects
## 
## formula: response ~ study + sex + partner_age
##             Df  logLik    AIC     LRT Pr(>Chi)  
## <none>         -342.41 700.82                   
## study        6 -336.36 700.71 12.1129   0.0595 .
## sex          3 -341.33 704.67  2.1546   0.5409  
## partner_age  3 -341.57 705.14  1.6823   0.6409  
## ---
## Signif. codes:  0 '***' 0.001 '**' 0.01 '*' 0.05 '.' 0.1 ' ' 1
\end{verbatim}
\end{kframe}
\end{knitrout}

or \texttt{scale\char`_test(clmodel)}, respectively. These tests can be viewed as goodness-of-fit tests. In the model above, there is no statistical evidence that including nominal effects into the model would increase the model fit significantly.

Finally, there are extensions to nonlinear ordinal regression models, i.e. model in which the predictors are included in a nonlinear way. We refer e.g. to the functions \texttt{ordglm} from the \texttt{gnlm} \cite{Swihart2019} package or \texttt{nordr} from the \texttt{gln} \cite{Turner2023} package for further details.
	
	\section{Conclusion}
	\label{sec:conclusion}
	
	The tools of the ordinal regression framework provide a method to appropriately analyse ordinally scaled data, particularly with only a few number of response categories. By applying ordinal regression we yield probabilities for each response category. This offers a more differentiated view, e.g., regarding proportions of participants consenting in contrast to considering just a mean score. Moreover, considering data on a latent scale, we have the possibility of testing group differences and marginal effects such as cumulative odds. If survey data is very finely graduated, such as for numeric rating scales or comprehensive quality-of-life questionnaires, the usage of ordinal methods is often limited as the estimated parameters are not that meaningful anymore in contrast to e.g. a mean or median score or, are even not feasible due to a too large number of parameters to be estimated. For the \textsc{Filippa} study, we demonstrated that a latent variable based ordinal regression analysis has the potential to identify factors influencing the willingness to study participation and to quantify the probability of consenting to the participation in fictional studies with differing levels of invasiveness given a range of demographic variables from the children and their legal representatives.

	\section*{Data availability statement}
	
	The raw data underlying the presented analyses will be made available by the authors upon reasonable request.
	
  	\section*{Acknowledgments}
  
  	We thank Thomas Asendorf for helpful discussions.
	
	\section*{Conflicts of interest and financial disclosures}
	
	The authors have no conflicts of interest to declare. We conducted this research with institutional resources only. Open access charges were funded by the Open Access Publication Funds of the Göttingen University, who had no role in study design, data collection and analysis, decision to publish, or preparation of the manuscript.
	
	\bibliography{filippa_tutorial.bib}

	\appendix

	\section{\texttt{ggplot} parameters}
	\label{sec:appendix}

\begin{knitrout}
\definecolor{shadecolor}{rgb}{0.969, 0.969, 0.969}\color{fgcolor}\begin{kframe}
\begin{alltt}
\hlstd{theme} \hlkwb{<-} \hlkwd{theme}\hlstd{(}\hlkwc{panel.background} \hlstd{=} \hlkwd{element_blank}\hlstd{(),}
        \hlkwc{panel.spacing} \hlstd{=} \hlkwd{unit}\hlstd{(}\hlnum{1}\hlstd{,} \hlstr{'lines'}\hlstd{),}
        \hlkwc{panel.border} \hlstd{=} \hlkwd{element_rect}\hlstd{(}\hlkwc{colour} \hlstd{=} \hlstr{'black'}\hlstd{,} \hlkwc{fill}\hlstd{=}\hlnum{NA}\hlstd{,} \hlkwc{linewidth}\hlstd{=}\hlnum{1}\hlstd{),}
        \hlkwc{axis.title.x} \hlstd{=} \hlkwd{element_text}\hlstd{(),}
        \hlkwc{axis.title.y} \hlstd{=} \hlkwd{element_text}\hlstd{(),}
        \hlkwc{axis.text.y} \hlstd{=} \hlkwd{element_text}\hlstd{(),}
        \hlkwc{axis.text.x} \hlstd{=} \hlkwd{element_text}\hlstd{(}\hlkwc{angle} \hlstd{=} \hlnum{30}\hlstd{,} \hlkwc{vjust}\hlstd{=}\hlnum{1}\hlstd{,} \hlkwc{hjust}\hlstd{=}\hlnum{1}\hlstd{),}
        \hlkwc{strip.text} \hlstd{=} \hlkwd{element_text}\hlstd{(),}
        \hlkwc{panel.grid.major.x} \hlstd{=} \hlkwd{element_blank}\hlstd{(),}
        \hlkwc{panel.grid.major.y} \hlstd{=} \hlkwd{element_line}\hlstd{(}\hlkwc{linewidth} \hlstd{=} \hlnum{0.5}\hlstd{,}
      \hlkwc{linetype} \hlstd{=} \hlstr{'dotted'}\hlstd{,} \hlkwc{colour} \hlstd{=} \hlstr{'lightgrey'}\hlstd{),}
        \hlkwc{panel.grid.minor.x} \hlstd{=} \hlkwd{element_blank}\hlstd{(),}
        \hlkwc{panel.grid.minor.y} \hlstd{=} \hlkwd{element_line}\hlstd{(}\hlkwc{linewidth} \hlstd{=} \hlnum{0.5}\hlstd{,}
    \hlkwc{linetype} \hlstd{=} \hlstr{'dotted'}\hlstd{,} \hlkwc{colour} \hlstd{=} \hlstr{'lightgrey'}\hlstd{)}
        \hlstd{)}
\hlstd{response_colours} \hlkwb{<-}\hlkwd{rep}\hlstd{(}\hlkwd{c}\hlstd{(}\hlstr{'#4DAF4A'}\hlstd{,} \hlstr{'#377EB8'}\hlstd{,} \hlstr{'#FFFF33'}\hlstd{,} \hlstr{'#FF7F00'}\hlstd{,} \hlstr{'#E41A1C'}\hlstd{),} \hlnum{6}\hlstd{)}
\end{alltt}
\end{kframe}
\end{knitrout}

\end{document}